\magnification=\magstep1
\vsize 23 true cm
\hsize 16 true cm
\def\\{\hfil\break}
\parindent 20pt
\parskip 8pt
\baselineskip 20pt
\centerline{{\bf SPHERICAL BLACK HOLES CANNOT SUPPORT SCALAR HAIR}}
\vskip 3 true cm
\centerline{D.Sudarsky$^{(1)}$ and T.Zannias$^{(1,2)}$\footnote{*}{NAFTA RES. 
FELLOW}}
\centerline{$^{(1)}$INSTITUTO CIENCIAS NUCLEARES}
\centerline{UNAM}
\centerline{Circuito Exterior, C.U.}
\centerline{A.Postal 70-543}
\centerline{04510 Mex. DF}
\centerline{ and}
\centerline{$^{(2)}$INST. de FISICA y MATEMATICAS}
\centerline{ UNIVERSIDAD  MICHOCANA S.N.H. }
\centerline{ Edificio C-3 }
\centerline{ Morelia Mich., Mex.}
\eject

\centerline{ {\bf ABSTRACT }}

The static spherically symmetric ``black hole solution" of the 
Einstein - conformally
invariant massless scalar field equations presented in
[1] and [2], is critically examined. It is shown 
 that the 
 stress energy tensor is ill-defined at the horizon, and that its
evaluation  through  suitable regularization  yields ambiguous
  results.
Consequently,  the configuration  fails to represent a 
genuine black hole
solution. With the removal of this  solution as a counterexample to 
 the no hair conjecture, we argue 
that the following  appears to be true: 

{{\bf Spherical black holes
cannot carry any kind of  classical scalar hair}}.

\vfill
\eject

The theory of a scalar field conformally coupled to  Einstein's gravity
 in four spacetime
 dimensions  is described by the action:
$$
S[g,\Phi] =\int \sqrt{-g} d^4 x [{1\over {16\pi G}} R- {1\over 12}\Phi^2 R +  
{1\over 2}\nabla_a \Phi\nabla^a \Phi]
$$
implying the following  equations of motion:
$$(1-{\alpha}{\Phi}^{2})R_{mn}={\alpha}[4\nabla_{m} \Phi
\nabla_{n}\Phi
-2 \Phi\nabla_{m}\nabla_{n}\Phi-
g_{mn}\nabla^{\l} \Phi\nabla_{\l}\Phi]\eqno(1a)$$
and
$$\nabla^{m}\nabla_{m}\Phi=0\eqno(1b)$$
 where all indices are four dimensional, 
$\alpha={k\over 6}$ with $k= 8\pi G$, and  $G$ stands for  
 the gravitational
coupling 
constant.

Using a suitable generation technique, Bekenstein [1] (see also [2]), 
constructed an exact asymptotically flat, static, spherical solution of the 
above equations.
Specifically, the  solution is 
described  by:

$$ds^{2}=-(1-{r_{o}\over r})^{2}dt^{2}+(1-{r_{o}\over r})^{-2}dr^{2}
+r^{2}d{\Omega}\eqno(2a)$$
$${\Phi}=-{C\over r-r_{o}}\eqno(2b)$$
where $r_{o}>0$  is  an arbitrary constant, and $C= C(r_0, {\alpha})
={r_{o}{\alpha}^{-{1\over 2}}}$.
The  coordinate  chart is  supposed to 
cover the 
domain of outer communications of a black hole spacetime with the 
the horizon located at $r=r_{o}$ 

The solution is interesting in several respects.
First, it explicitly demonstrates that a given spacetime geometry 
 may locally
support two entirely different  stress tensor as sources of the 
Einstein equations. 
In the present case, the 
 geometry is that of the extreme Reissner-Nordstrom spacetime
 and  corresponds to a solution of the Einstein's equations with a source
given,  either by
 the stress
tensor of a $U(1)$ field, or the stress tensor of a conformal massless
scalar field.
Furthermore,  the solution (2-ab) is unique in the following sense: 
All asymptotically flat, static, 
spherically symmetric solutions 
of (1a,b) with nontrivial $\Phi$, other than the Bekenstein solution (2a,b),  
do not posses
 a regular horizon [3].
 
Finally, the scalar field ${\Phi}$  given in  (2b) is diverging  on the 
horizon. Initially, this  divergence was considered as a pathology of 
the solution, and a black hole interpretation of the solution
was not  advanced. However, further analysis in [4], suggested
that the divergence of ${\Phi}$ on the 
horizon may be innocuous. The regularity of 
the geometry at the horizon, ensures that tidal gravitational forces are
bounded there.
Test particles following geodesics, 
``feel" nothing peculiar as they cross the horizon, since, as we 
have already mentioned,
the spacetime geometry is that of the extreme Reissner-Nordstrom 
black hole which possess  a regular horizon. It was, moreover, argued that,
even when considering particles coupled to the scalar field itself,  
physical pathologies 
 due to the divergence were not expected,
 as these particles would require an 
infinite
amount of proper time to reach the horizon [4]. 
Consequently, Bekenstein, with the  encouragement of DeWitt [4],
reconsidered previous thoughts [1] and interpreted 
(2a,b) as  genuine 
 family of black hole solutions of
Einstein's- conformally invariant massless scalar field equations.
Partially, however, due to the fact that the solution appears to be 
unstable [5], and partially due to the  lack of evidence 
of long 
range forces mediated by  scalar field, astrophysically, 
the solution never enjoyed
the same status as the well known family of Kerr-Newman black holes.
It does, however, raise the following question: One of the basic requirements
used in 
the establishment of black hole uniqueness theorems is that  
of a regular event horizon. Whenever scalar fields are involved, regularity
of the horizon is 
normally associated with  the requirement that 
the scalar field is  bounded  at the horizon. If, on the other
hand, the Bekenstein solution is a genuine black hole, then one would 
need to reexamine the 
established uniqueness theorems
by  relaxing  the boundedness  behavior of the scalar fields on the 
horizon. It was this issue that motivated us to take a closer look
at the Bekenstein solution.

Another interesting aspect of (2a-b) is its relation to  the 
no hair conjecture.
Let us note that the  new family is  
characterized by the $ADM$ mass ${r_{o}\over 2}$ and scalar charge
$Q=r_{o}{\alpha}^{-{1\over 2}}$. On the other 
hand,  only the $ADM$ mass defined 
 by a two- surface integral at infinity, is associated with an asymptotic
conservation law.
The same parameter characterizes also the familiar Schwarzschild
black holes; therefore, 
the solution (2-ab) carries ``hair'' and indeed represents a counterexample
to Wheeler's graphical statement `` black holes have no hair'' a
precursor of the  "No Hair Conjecture". The ``No Hair Conjecture''
has often been interpreted in a different way by different authors, so
we shall be more precise.
Following Bizon [6], we would say that a certain theory 
allows a hairy black hole if 
there is a need to
specify quantities other than conserved charges defined at asymptotic infinity, 
in order to
characterize completely a stationary black hole solution within that theory.
With this definition, the well known dilatonic black hole solution [7, 8] 
does not constitute a
hairy solution, by virtue  of the fact that it is the unique  static
 solution [9] of that theory, and  it is specified  completely by the
 values of the electric charge and of its  ADM mass ( the Reissner- 
Nordstrom  configuration is not  
a solution of the theory).
 
Having clarified the meaning of ``hair'',
we shall argue in the present paper against a black hole interpretation of 
the solution (2-ab). In fact, we shall show that the divergence of the 
field ${\Phi}$
at the horizon has rather severe consequences.
Specifically, we shall show that the extension of (2-ab) 
to a chart that  includes the horizon, fails to satisfy Einstein's equation 
at the horizon,
and,  therefore, it cannot be considered as a genuine black hole solution.

To do so, let us first rewrite (2) in advanced Eddington-Finkelstein
coordinates. Such coordinate chart is advantageous
since it can be extended through the horizon 
(in fact a portion of it). One then has:

$$ds^{2}=-V^{2}du^{2}+2dudr+r^{2}d{\Omega}\eqno(3)$$ 
 
where $V^{2}=-{\xi}^{a}{\xi}_{a}=(1-{r_{o}\over r})^{2}$ is the 
square of the Killing field that is timelike at infinity,
and $r$ is a  null coordinate varying in  the range $ (0, \infty)$.
 The  extension of the metric is given by the same expression in all the chart, 
 but  the extension of the scalar field through the horizon has 
 a sign ambiguity.
This is easily resolved by noting that, in these coordinates, the field equation
for $\Phi=\Phi(r)$ admits  a first integral:
$$
{V^{2} r^{2}} {\partial{\Phi}\over {\partial r}}=const \eqno(4)
$$
thus,  the expression for the scalar field is given by eq.( 2.b) 
throughout the chart (in fact, we will see that we have to give 
a distributional meaning to this configuration, and that only 
this extension can be considered as a
distributional solution).

In view of the fact that the metric  is regular at $r=r_{o}$,
we
 shall  explicitly  check whether or not Einstein's equations 
hold at the horizon. The fact  that
 $l^a =({{\partial}\over{\partial r}})^a $ 
is a  smooth null vector field (even
across $r=r_{o}$), implies that  the quantity $R_{mn}l^ml^n $ is 
finite everywhere,
 and, in particular, at the horizon. Moreover, direct 
calculation  shows that it 
actually vanishes. The next  step would be to  compute
$T_{mn}l^{m}l^{n}|_{r_{o}}$ and  compare its value with that of 
$R_{mn}l^ml^n|_{r_{o}}$.
However, due to the unboundedness of ${\Phi}$, 
special care must be taken.
One may naively  compute  $T_{mn}l^{m}l^{n}|_{r_{o}}$
 by
identifying it with
$$lim_{r \to r_0} (T_{mn}l^{m}l^{n}|_{r})\eqno(5)$$
Computing explicitly the right-hand side of (5) for $r\not = r_0$ leads to
$$
T_{mn}l^{m}l^{n}|_{r}= {{\alpha} \over {(1-{\alpha}{\Phi}^{2})}}
[4({\Phi}')^{2}-2{\Phi}{\Phi}'']=O\eqno(6)$$
where the prime indicates derivative with respect to $r$. We thus find that
 the limit is well-defined and is zero. 
This identification would suggest that Einstein's equations hold
at the horizon. However, that is not correct. 
The problem  
with the above procedure is that this result follows from the fact that 
Einstein's equations hold everywhere outside the horizon, and thus they
obviously hold in the 
limit as we approach the
horizon. 
However, this limiting procedure,  still  {\it does not  
tell us  whether or not Einstein's equations  hold at the horizon}.
In other words, we still do not know what the  value 
of $T_{mn}l^{m}l^{n}|_{r_{o}}$ actually is.
If indeed  $T_{mn}l^{m}l^{n}|_{r}$ was continuous  in a neighborhood 
of $r_{o}$, one could identify the value of $T_{mn}l^{m}l^{n}|_{r_{o}}$ 
with the  limit obtained above, but that is precisely the 
question we are addressing. In fact, the right-hand side of (6)
at $r_0$ is of the form
$ 0\times(\infty -\infty)$, and thus  is ill-defined.
 
 The problem, of course, has to do with the fact that the scalar field 
is not really 
well-defined throughout the spacetime, and, in particular, at the horizon. In
order to try to make sense of this field configuration, we might consider the
generalized solutions, i.e., solutions in the distributional sense, and 
in particular we will take $\Phi$
 as the principal value  of ${-C\over {r-r_0}}$  distribution .
This means considering the solution as a functional on the space of test functions
 $X =C_0^\infty (M)$ of infinitely differentiable
 functions of compact support in the manifold $M$, and to define for all $f\in X$
$$
<\phi,f> = lim_{\epsilon \to 0} \lbrace
\int_{D^1(\epsilon)} {-C\over {r-r_0}} f \sqrt{|g|} d^4x + 
\int_{D^2(\epsilon)} {-C\over {r-r_0}} f \sqrt{|g|} d^4x \rbrace.
\eqno(7)
$$ 
where $D^1 (\epsilon) =\lbrace x\in M / r(x) >r_0 +\epsilon\rbrace$ and
 $D^2(\epsilon )=\lbrace x\in M / r(x) >r_0 +\epsilon\rbrace$.
 We then ask whether or not this $\phi$ is a distributional solution of
the scalar field equation (1b).
 This 
 question is then whether, for all $f\in X$,
$$
<\nabla^{m}\nabla_{m}\Phi,f> \equiv <\Phi,\nabla^{m}\nabla_{m}f>=0
\eqno(8)
$$
 The answer is in the affirmative, has been seen from the following evaluation:
$$
 <\Phi,\nabla^{m}\nabla_{m}f>=lim_{\epsilon \to 0} \lbrace
\int_{D^1_\epsilon} {-C\over {r-r_0}} \nabla^{m}\nabla_{m}f \sqrt{|g|} d^4x + 
\int_{D^2_\epsilon }{-C\over {r-r_0}} \nabla^{m}\nabla_{m}f \sqrt{|g|} d^4x \rbrace.
\eqno(9)
$$
integrating by parts, and recalling that $f$ has compact support, we find

$$
 <\Phi,\nabla^{m}\nabla_{m}f>= -C \times lim_{\epsilon \to 0} \lbrace
\int_{\partial D^1 (\epsilon)  }
[ {1\over {r-r_0}} \nabla_{m}f - f \nabla_{m}{1\over {r-r_0}}]
 n^{m}\sqrt{|h|} d^3x 
$$
$$
 + \int_{\partial D^2 (\epsilon) }
[ {1\over {r-r_0}} \nabla_{m}f - f \nabla_{m}{1\over {r-r_0}}]
 n^{m}\sqrt{|h|} d^3x \rbrace.
\eqno(10)
$$
where $\partial D$ indicates the boundary of the region $D$,
 $h$ is the determinant of the induced metric on the boundary, and
 $n^m$ is the outward pointing  unit normal to it. Thus
$$
 <\Phi,\nabla^{m}\nabla_{m}f>=-C \times  lim_{\epsilon \to 0} \lbrace
\int_{r=r_0+\epsilon} [ (1/\epsilon) {\partial f\over {\partial u}}
+{\epsilon\over r^2} {\partial f\over {\partial r}} +{1\over r^2}f]
r^2 \sin(\theta) d \theta d\phi du
$$
$$ 
-\int_{r=r_0-\epsilon} [ (-1/\epsilon) {\partial f\over {\partial u}}
+{\epsilon\over r^2} {\partial f\over {\partial r}} +{1\over r^2}f]
r^2 \sin(\theta) d \theta d\phi du \rbrace
\eqno(11)
$$
where we have used $ n^m= (1-r_0/r)^{-1} (\partial /\partial u)^m +
(1-r_0/r) (\partial /\partial r)^m$ and 
$\sqrt{|h|} =|(1-r_0/r)| r^2  \sin(\theta)$. 
Noting that the first term in the integrals can be integrated out to
yield zero, because $f$ has compact support, and 
that the remaining integrals
cancel out, due to the continuity of $f$ and its derivatives, we see that, in 
effect, we have a distributional solution to the scalar field equation.
That this is not a trivial result is evidenced by the well known fact that
$1/r$ is a solution of sourceless Laplace's equation in in flat $R^3 -{0}$ but that 
when the 
origin is included we have a distributional solution
 of Laplace's equation with a $\delta$ "funtion" source.
Moreover the above calculation shows that we had to take the current extension of the scalar
field through the horizon in order to have a distributional solution.

Next, we turn to  Einstein's equations. We note that the spacetime metric, 
being the same as that in the Reissner Nordstrom solution, 
is  regular everywhere,
so the right-hand side of  Einstein's equations presents no problem. The
issue is then the left-hand side of the equations, namely the 
energy momentum tensor. The first issue is whether we can
 give any meaning to it.
The problem is similar to the one often found in quantum field theory, 
and that is the origin of the infinities that plague
the theory, i.e., the fact that one is forced to deal with expressions
that contain products of distributions at ``the same point".
The latter is an operation that is not mathematically well-defined.
In quantum field theory, this problem is dealt with through the process of
 regularization and the subsequent renormalization of the expressions
through
the subtraction of formally divergent terms in a well-defined fashion.
In the particular case of the energy momentum operator in quantum
 field theory in
curved spacetime, the renormalization consists in the subtraction of 
divergent terms corresponding to the  vacuum expectation value of the
tensor operator in Minkowski spacetime. More precisely, the Hadamard
anzats for the bi- distribution used in the substraction scheme that 
renormalizes the energy- momentum tensor is motivated by the vaccum 
two point funtion in Minkowski spacetime [10].
In our case, since we are
 dealing with classical solutions, the
analogous terms would correspond to the energy momentum tensor for
 the configuration
$\Phi=0$ in Minkowski spacetime, which is $T_{mn}=0$; thus, we do not have any
 canonical expression that can be subtracted in order to renormalize
 the ill-defined quantities we are dealing with.

We could consider whether it is possible to give meaning to the expression 
in question by means of a regularization without the subsequent renormalization.
The fact that the product of distributions is not well-defined suggests the answer 
is negative, but let's examine specifically the problems we encounter
 in the attempts to do so.

 We start by noting that the theory of distributions provides the means to 
regularize expressions through the following theorem [11]:

{\it Every distribution is the limit in the distributional sense 
of functions of class $C^\infty_0$}.

The idea is, then, to try to assign a value, to $T_{mn}l^{m}l^{n}|_{r}$
through a regularization procedure. In fact it is more convenient to
look at both sides of eq. (1a) contracted with $l^m l^n$ and to consider 
their regularized values. To this effect 
we consider ${\Phi}$ as the (distributional) limit when $a\to 0$
of a class of $C^\infty_0$ functions ${\Phi}_{a}$.  We must
chose ${\Phi}_{a}$  so that it tends to $\Phi$ pointwise everywhere, except at
 the horizon (because, when considering test functions with support away from
the horizon, the distribution $\Phi$
corresponds to a regular function).

Next, we  note that the RHS. of eq (1a) vanishes identically
(because the metric  (3) has $R_{mn} l^m l^n =0$) and finally we turn to the LHS
of eq (1a) contracted with  $l^m l^n$ 
which is given by
 $A(\Phi) \equiv {2 \alpha}
[2({\Phi}')^{2}-{\Phi}{\Phi}''] $   To actually compute it we  
 replace $\Phi$ by  
its regularized version ${\Phi}_{a}$, 
and, at the end, remove the regulator (i.e., take  the limit $a \to 0$). 
In that manner, if 
 the R.H.S. is well-defined,  the procedure would yield a
well-defined distribution. Moreover, if the energy momentum
 tensor is to be considered
as well-behaved, the procedure should yield a finite expression; 
and if, furthermore, Einstein's equations are to be said to hold, the resulting 
value  should be the zero distribution. 

We can employ as   ${\Phi}_{1,a}$ the function
given by [12]:
$$
{\Phi}_{1,a}={C(r-r_{o})\over a^2 +(r-r_{o})^{2}}
\eqno(12)
$$ 
One may easily verify that as $a \to 0$, $ {\Phi}_{1,a} $
 converges pointwise to $\Phi$ everywhere in the Eddington- Finkelstein chart
minus the horizon.

Carrying through the above procedure, we obtain:
$$
A(\Phi_{1,a})={4 r_{o}^2\over a^{4}}{1\over{ (1+y^{2})^3
}} \eqno(13)
$$
where $y={r-r_{0}\over a}$.
Therefore, at the horizon, one finds  
$A= {4r_{o}^2\over a^{4}}$. 
This seems to indicate that the R.H.S of eq. (1,a)
is  divergent at the horizon with a
singularity of
the type   $(\delta)^{4}$ ( The hight of A is order $a^{-4}$ and its width is of
 order $a$).

However, we can take a different class of regularizing functions [13],
 namely:

$$
{\Phi}_{2,a}={C(r-r_{o})^3\over (a^2+(r-r_{o})^{2})^2}
\eqno(14)
$$ 
or more  generally

$$
{\Phi}_{n,a}={C(r-r_{o})^{2n-1}\over (a^2+(r-r_{o})^{2})^{2n}}
\eqno(15)
$$
These classes of functions also
 converge pointwise to $\Phi$ as $a \to 0$ everywhere 
in the Eddington- Finkelstein chart
minus the horizon.

Now, if we use the functions ${\Phi}_{2,a}$, and repeat the procedure we find:
$$
A(\Phi_{2,a})={8 r_{o}^2\over a^{4}}{y^4 (3+y^2)\over{ (1+y^{2})^6}} 
\eqno(16)
$$
where $ y ={r-r_{0}\over a}$. Evaluating 
at $r-r_{0}$, we find $A(\Phi_{2,a}) (y=0)=0$,
in contrast with what was found using the first regularization.
In this case, we actually have a divergence, but of a more complicated nature,
in fact the form of eq. (16) suggests a singularity of the type $(\delta''- \delta)^4$.
The main point is, however, that the result depends on the form of the functions
employed in the regularization, which indicates the type of problems
 one encounters in trying to 
give meaning to the product of singular distributions. In other words, we see, 
in a clear way, that there is no canonical  procedure to give meaning to
the R.H.S of eq. (1.a) . This would be, however, an ovious prerequisite
that must be satisfied in order
 to be able to state that one has a solution of Einstein's equations.

The above discussion shows that  Bekenstein's configuration cannot
 be considered
 to satisfy the Einstein's equations on the whole manifold including
the  horizon. Consequently, it cannot represent a  
regular black hole spacetime solution.
On the other hand, the scalar field equation can
be considered as satisfied in the distributional sense.
And, of course, for $r>r_0$, we have  a perfectly valid solution
of the coupled Einstein-scalar field equations.

 The removal of the Bekenstein solution as a possible black hole
 changes the scenario describing  black holes 
admitting 
classical scalar hair.
In view of the fact  that
spherical black holes cannot carry hair of massive - massless [14], or even 
arbitrarily self interacting scalar field, as long as it is minimally
coupled to gravity [15, 16], combined with the results of [17]
for the  conformal coupling and those of [18]
for arbitrary coupling, appears to indicate strongly the 
validity of the following 
conjecture:

{\bf Spherical black holes
cannot carry any kind of classical hair associated to a scalar field}.

On physical grounds, one may attribute the absence of any kind of 
scalar hair as due to the fact that scalar mediated forces are attractive in 
nature, and, thus, there is nothing to counterbalance the equally 
attractive  gravitational
force.
In contrast,  vector mediated forces can be repulsive, and, thus, help to
balance the
gravitational attraction [19]. Needles to say that all those issues need 
a closer 
examination.

\centerline{ {\bf ACKNOWLEDGEMENTS }}

 We wish to thank Rafael Sorkin for very helpful discussions that led to the 
present analysis. D.S. acknowledges partial support from the DGAPA-UNAM
 project IN-105496 and
T.Z. acknowledges  partial support through a NAFTA-NSERC fellowship that made
the visit 
to  ICN-UNAM possible.  T.Z. also is grateful for the  warm hospitality 
at ICN-UNAM, where part of this work has been done.
\eject

\centerline{ {\bf REFERENCES }}

\\
[1] J.D.Bekenstein Ann.Phys.(N.Y) 82, 535 (1974)\\
Actually Bekenstein's original solution included also an electromagnetic field.\\
However, for simplicity in the present paper, its contribution will be ignored.\\
Its inclusion does not change the issue at hand.\\
[2] N. Bocharova, K. Bronikov and V. Melnikov, Vestn. Mosk. Univ.\\
 Fiz. Astron. 6, 706, (1970) \\
[3] B. C. Xanthopoulos and T. Zannias J.Math.Phys.32,1875,(1991)\\
[4] J.D.Bekenstein Ann.Phys.(N.Y.) 91,72. (1975)\\ 
[5] K. Bronikov and K. Kireyev Phys. Lett. A67, 95, (1978)\\
[6] P. Bizon preprint, Jagellonian Univ. Instit. of Phys. Cracow Pol.(1994)\\
[7] G. W. Gibbons Nucl.Phys. B207,337,(1982)\\ 
[8] D. Garfinkle, G. Horowitz, A. Strominger Phys.Rew.D43,3140 (1991)\\
(Err.D45,3888,1992)\\
[9] A. K. M. Masood-ul-Alam Class. Quantum Grav. 10, 2649, (1993)\\
[10] See for example section 4.6 of R. M. Wald,
 Quantum Field Theory in Curved Spacetime and Black Hole Thermodynamics,
The University of Chicago Press (1994).\\
[11] See for example Theorem 5.2.2  F.G. Friedlander, Introduction to the theory
of Distributions, Cambridge University Press (1982).\\
[12] The regularization scheme employed is rather a standard one.\\
See for example J. D. Jackson, Classical Electrodynamics, Wiley N.Y.(1975)\\
 for an elementary example, or for a more complicated,  case see:\\ 
J. Louko and R. Sorkin, Class. Quant. Grav. (in press).\\
[13] We thank an anonymous referee for pointing out this  new regularization.\\
[14] J. D .Bekenstein, Phys. Rev. D5,2941 (1972)\\
 J. E. Chase, Commun. Math. Phys. 19,276 (1970)\\   
 A. Mayo and J. Bekenstein Phys. Rev. D54, 5059, (1996)\\
[15] M. Heusler, J. Math. Phys. 33, 3497, (1992) \\
[16] D. Sudarsky, Class. Quant. Grav. 12, 579, (1995) \\
[17] T. Zannias J. Math. Phys. 36, 6970 (1995)\\
[18] A. Saa  Phys. Rev. D53, 7373 (1996) and J. Math. Phys. 37, 2349, (1996)\\
[19] G. Gibbons Private communication\\

\bye